\documentclass[conference]{IEEEtran}
\IEEEoverridecommandlockouts
\usepackage{cite}
\usepackage{amsmath,amssymb,amsfonts}
\usepackage{algorithmic}
\usepackage{graphicx}
\usepackage{textcomp}
\usepackage{xcolor}
\usepackage[algo2e,linesnumbered,ruled]{algorithm2e}

\newtheorem{definition}{Definition}

\def\BibTeX{{\rm B\kern-.05em{\sc i\kern-.025em b}\kern-.08em
		T\kern-.1667em\lower.7ex\hbox{E}\kern-.125emX}}
\begin{document}

\title{A Predictive On-Demand Placement of UAV Base Stations Using Echo State Network
}

\author{
	\IEEEauthorblockN{Haoran Peng\IEEEauthorrefmark{1}, Chao Chen\IEEEauthorrefmark{2}, Chuan-Chi Lai\IEEEauthorrefmark{1}, Li-Chun Wang\IEEEauthorrefmark{1}, and Zhu Han\IEEEauthorrefmark{3}}
	\IEEEauthorblockA{\IEEEauthorrefmark{1}Department of Electrical and Computer Engineering, National Chiao Tung University, Hsinchu 300, Taiwan\\
	\IEEEauthorrefmark{2}Institute of Network Engineering, National Chiao Tung University, Hsinchu 300, Taiwan\\
	\IEEEauthorrefmark{3}Department of Electrical and Computer Engineering, University of Houston, Houston, TX 77004, USA\\
	}	
	
}


%


\maketitle

\begin{abstract}
The unmanned aerial vehicles base stations (UAV-BSs) have great potential in being widely used in many dynamic application scenarios. In those scenarios, the movements of served user equipments (UEs) are inevitable, so the UAV-BSs needs to be re-positioned dynamically for providing seamless services. In this paper, we propose a system framework consisting of UEs clustering, UAV-BS placement, UEs trajectories prediction, and UAV-BS reposition matching scheme, to serve the UEs seamlessly as well as minimize the energy cost of UAV-BSs' reposition trajectories. An Echo State Network (ESN) based algorithm for predicting the future trajectories of UEs and a Kuhn-Munkres-based algorithm for finding the energy-efficient reposition trajectories of UAV-BSs is designed, respectively. We conduct a simulation using a real open dataset for performance validation. The simulation results indicate that the proposed framework achieves high prediction accuracy and provides the energy-efficient matching scheme.
\end{abstract}


%
\IEEEpeerreviewmaketitle

\section{Introduction}
\label{intro}
With the rapid development of cellular communication techniques, people become dependent on mobile devices, such as cell phones and tablets in daily life. Most communication services for these devices are provided by ground base stations (BSs). However, in some remote or hot-spot areas where ground BSs are unable to cover, the quality of communications service is usually very poor. To solve this problem, the use of unmanned aerial vehicles (UAVs) as BSs has been a popular research topic in recent years. UAV-BSs can deliver reliable, cost-effective, and on-demand cellular communication service especially to dynamic scenarios~\cite{DBLP:journals/corr/abs-1803-00680}.

Recent study~\cite{7848883} on UAV-BSs mainly focuses on finding the optimal placement of the UAV-BSs while serving the user equipments (UEs) in the target area. However, in most papers~\cite{7918510}~\cite{7881122}~\cite{8642333}, few take dynamic scenarios into consideration. In real scenarios, the UAV-BSs are expected to dynamically reposition in response to the dynamic movement of UEs. In~\cite{7417609}, the approach considered dynamic scenarios, but the reposition action is performed after the movement of UEs, which may cause communication service outage. In order to serve the UEs seamlessly, a prediction on the future positions of UEs can be performed before the reposition action of UAV-BSs in the next time slot. This prediction can be implemented by applying an \emph{Echo State Network} (ESN), a type of recurrent neural network.
\begin{figure}[ht]
	\centering
	\includegraphics[width=0.45\textwidth]{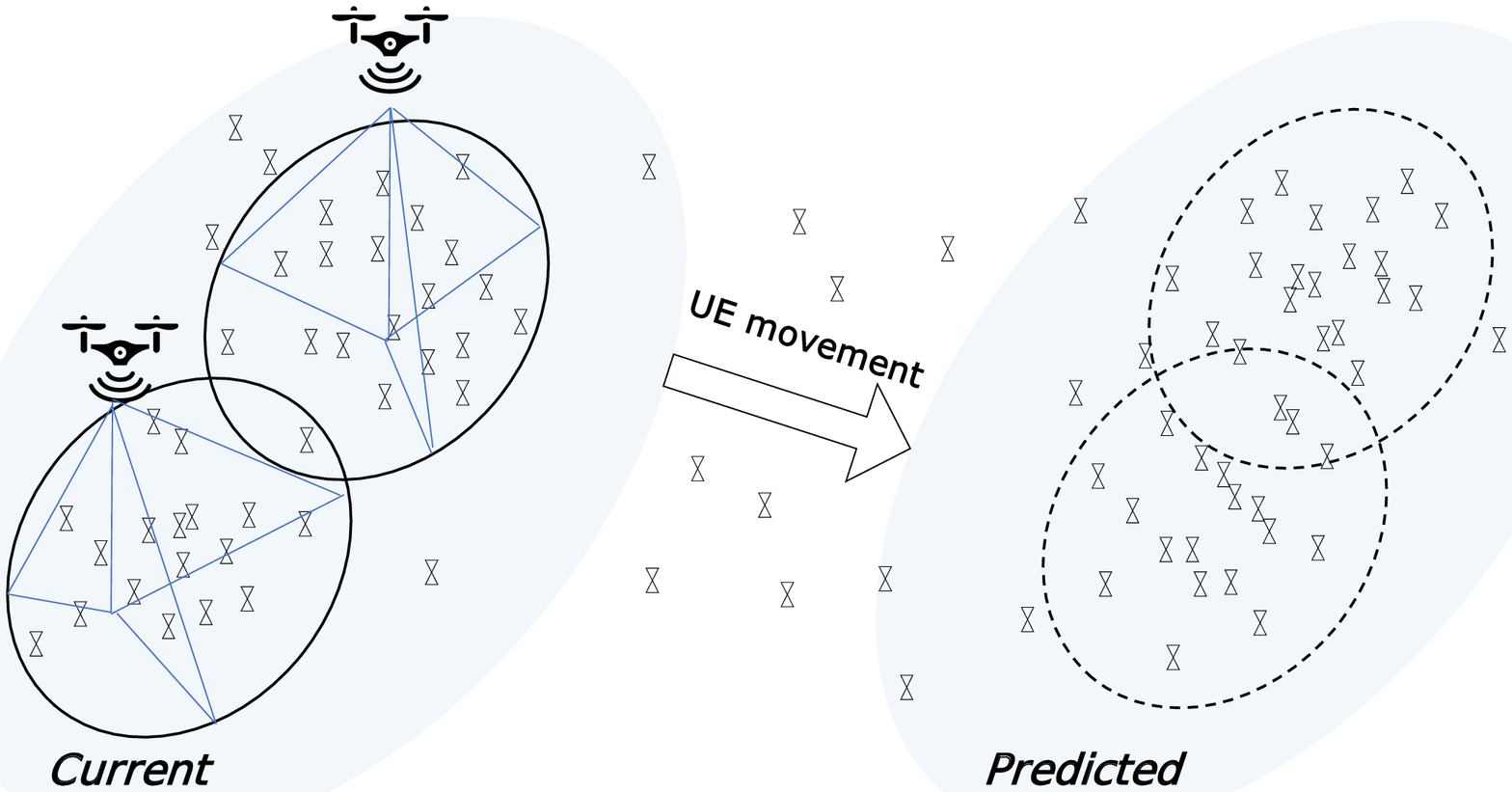}
	\caption{The considered dynamic application scenario.}
	\label{fig:system}
\end{figure}

The main contribution of this paper is to propose a system framework for dynamic placement of multiple UAV-BSs, aiming to serve the UEs seamlessly and minimize the energy cost of UAV-BSs' trajectories. The merits of this work are described from the following three aspects:
\begin{itemize}
	\item We design an ESN-based algorithm for predicting the future trajectories of UEs, aiming at serving the UEs seamlessly. With the precise predicted locations of UEs, the system can search for the potential locations of UAV-BSs in the next time slot for the seamless services.
	\item We develop a Kuhn-Munkres-based weighted bipartite matching algorithm to find the energy-efficient trajectories of multiple UAV-BSs when reposition from current positions to the predicted positions in the next time slot. This algorithm is aiming at minimizing the energy cost in the reposition action phase.
	\item We adopt the GeoLife dataset~\cite{DBLP:journals/debu/ZhengXM10} as the input information with some preprocessing steps for our simulations. The simulation results indicate that the proposed algorithm can achieve high accuracy in terms of root mean square error (RMSE).
\end{itemize}

The remaining parts of this paper are organized as follows. In Section~\ref{sec:related_work}, the related work is introduced. In Section~\ref{sec:framework}, we discuss the ESN-based prediction algorithm, the KM-based matching algorithm, and the proposed algorithm step by step. In Section~\ref{sec:simulation}, the simulation results are discussed in detail. We make the conclusion remarks in Section~\ref{sec:conclusion}.

\section{Related Work}
\label{sec:related_work}
For UAV control, in~\cite{7417609}, the authors studied on finding the optimal altitude for a single UAV-BS with the maximum coverage and minimum required transmit power. The authors in~\cite{7918510} showed that the UAV-BS placement in the horizontal dimension can be modeled as a circle placement problem and a smallest enclosing circle problem, and they proposed an optimal 3D placement algorithm for maximizing the number of served users with the minimum transmit power. In~\cite{7881122}, a method using the heuristic algorithm for finding the positions of UAV-BSs in an area with different user densities was proposed. This algorithm can estimate the minimum number of UAV-BSs and their placement, while satisfying coverage and capacity constraints. In~\cite{8642333}, the placement problem was modeled as a knapsack-like problem, and a density-aware placement algorithm to maximize the number of covered users with the constraint of the minimum required data rates per user was proposed. 

For learning, the authors in~\cite{8445768} proposed a Q-learning based method for finding the optimal trajectory of an UAV-BS to serve multiple users. In their method, the UAV-BS acts as an autonomous agent to learn the trajectory that maximizes the sum rate of the transmission. A dataset from a project called GeoLife is introduced in~\cite{DBLP:journals/debu/ZhengXM10}. This dataset consists of 182 users' locations and GPS trajectories in a period of over five years. In~\cite{8455627}, the authors proposed an Markov Chain predictive model for predicting the next location of an individual based on its recent locations and mobility behavior over a period of time. In their simulation part, the GeoLife dataset was used to evaluate the proposed algorithm.

\section{Proposed System Framework}
\label{sec:framework}
We consider the dynamic application scenario depicted in Fig.~\ref{fig:system}. The UAV-BSs are expected to reposition to the potential positions in advance of the movement of the UEs. In Section~\ref{sec:esn}, we design an ESN-based algorithm to predict the future trajectories of UEs for finding the potential positions of UAV-BSs. In Section~\ref{sec:km}, we also take the energy cost of the reposition action of UAV-BSs into account. A KM-based matching algorithm is proposed to find the energy-efficient trajectories. In Section~\ref{sec:algo}, the proposed algorithm consisting of clustering, placement, prediction, and matching is introduced step by step.

\subsection{ESN-based Prediction Algorithm}
\label{sec:esn}
To obtain the predicted positions of UEs in the next time slot, an algorithm based on ESN is proposed. We choose ESN due to its low computation time and energy cost.
ESN forms a hidden layer of the network by randomly deploying a large numbers of neurons, known as the reservoir pool. The ESN model has the following characteristics:
\begin{enumerate}
	\item containing a large number of neurons;
	\item the connection between neurons is generated randomly; 	
	\item the links between neurons are sparsity.
\end{enumerate}

\begin{figure}[t]
	\centering
	\includegraphics[width=0.4\textwidth]{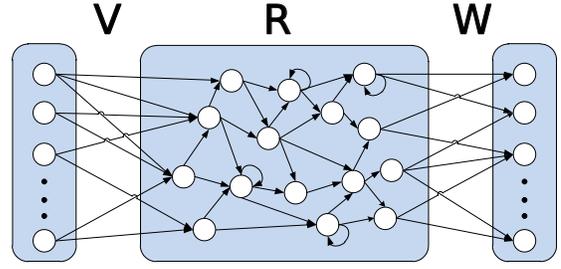}
	\caption{The framework of an echo state network.}	
	\label{fig:esn}
\end{figure}

The structure of the ESN model is drawn in Fig.~\ref{fig:esn}, where $V$ represents the input weight matrix, $R$ is the reservoir weight matrix and $W$ is the output weight matrix.

In the input layer, we define the input vector as $\mathbf{u}^{n\times 1}$, and its dimensions is $n\times 1$. In the reservoir pool, the typical update equation is written as~\cite{2001_Jaeger}
\begin{equation}\label{eq:esn:update}
\mathbf{x}^{m\times 1}(n)=\tanh(\mathbf{W}_{in}^{m\times n}\mathbf{u}^{n\times 1}(n)+\mathbf{W}_{x}^{m\times 1}(n-1)),
\end{equation}
where $\mathbf{x}^{m\times 1}$ is a vector of internal units in the reservoir pool. $\mathbf{W}_{in}^{m\times n}$ is the connection weight matrices between input layer and reservoir pool, and $\mathbf{W}_x^{m\times 1}$ is the recurrent weight matrices. 

The reservoir pool is linearly connected to the output layer, which can be defined as
\begin{equation}\label{eq:esn:reservoir:pool}
\mathbf{y}(n)=\mathbf{W}_{out}^{n\times m}\mathbf{x}^{m\times 1}(n),
\end{equation}
where $\mathbf{y}(n)$ is the output vector and $\mathbf{W}_{out}^{n\times m}$ is the connection weight matrices between reservoir pool and output layer.

We use root mean square error (RMSE) to evaluate the quality of this model, the expected value is $\hat{\mathbf{y}}_i$ and the actual result is $\mathbf{y}_i$. RMSE is defined as
\begin{equation}\label{eq:esn:mse}
\text{RMSE}=\sqrt{\dfrac{1}{M}\sum_{i=1}^{M}w_i(\mathbf{y}_i-\hat{\mathbf{y}}_i)^2},
\end{equation}
where $\sum_{i=1}^{M}w_i=1$, $M$ is the number of predictions, and $w_i$ are the weights. In the considered dataset, some UEs move unexpectedly (U-turn, right-angled turn), and the positions of these unexpected movements are assigned with the smaller weights.

%

\subsection{KM-based Matching Algorithm}
\label{sec:km}
\begin{figure}[!ht]
	\centering
	\includegraphics[width=0.25\textwidth]{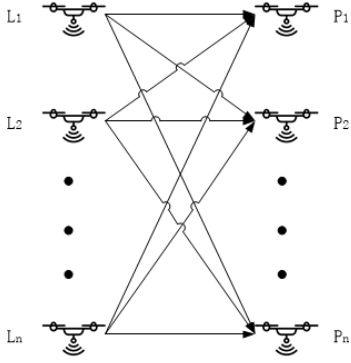}
	\caption{Reposition matching of multiple UAV-BSs.}	
	\label{fig:MWBMP}
\end{figure}

Without considering environmental factors, if the summation of all UAV-BSs' moving distance from the current positions to the positions in the next time slot is smaller, the less energy will be consumed. Such an \emph{Energy-Efficient Displacement Optimization} (EEDO) problem can be formulated as Definition~\ref{def:problem:EEDO}. The energy-efficient placement optimization problem also can be reduced from a well-known problem, \emph{Minimum Weighted Perfect Bipartite Matching}, as shown in Fig.~\ref{fig:MWBMP}.

\begin{definition}[EEDO]\label{def:problem:EEDO}
	At time $t$, the current position of a UAV is denoted as $L_i(t)\in \mathbb{R}^2$, and the potential position of a deployed UAV in the next time slot is denoted as $P_j(t)\in \mathbb{R}^2, i,j\in[1,n]$, $n$ is the number of UAVs, and $w_{i,j}$ is the energy cost for a UAV to fly from position $L_i$ to position $P_j$. The total energy cost of the UAV-BSs' displacements can be optimized by
	\begin{align}\label{eq:cost}
	\min&\sum_{\forall i,j\in[1,n]}w_{i,j}x_{i,j}&\tag{P1}\\
	\text{s.t.}&\hspace{.5em}\sum_{i=1}^{n}x_{i,j}=1,& \forall j=1,2,\dots,n,\nonumber\\
	&\hspace{.5em}\sum_{j=1}^{n}x_{i,j}=1,& \forall i=1,2,\dots,n,\nonumber\\
	&\hspace{.5em}x_{i,j}\in\{0,1\}.& \nonumber
	\end{align}
\end{definition}


In this work, we assume the number of UAV-BSs is $n$, and the heights of the multiple UAV-BSs are the same and equal to a constant. At time $t$, the number of current positions and the number of predicted positions of UAV-BSs are both equal to $n$. The coordinate of the $i$th current position is defined as $L_i(t)=[x_i(t),y_i(t)]^T\in\mathbb{R}^2$ and the coordinate of the $j$th predicated position is defined as $P_j(t)=[\bar{x_j}(t),\bar{y_j}(t)]^T\in\mathbb{R}^2,\forall i,j\in[1,n]$, where $T$ is a predefined time period slot between the predictions, and $x_i(t)$ and $y_i(t)$ are the latitude and longitude of the $i$th current position, respectively. $\bar{x_j}(t)$ and $\bar{y_j}(t)$ are the latitude and longitude of the $j$th predicated position, respectively. The vertex labeling of $L_i(t)$ and $P_j(t)$ are denoted as $A[L_i(t)]$ and $B[P_j(t)]$, respectively. The weight between $L_i(t)$ and $P_j(t)$ is defined as $[L_i(t)][P_j(t)]$, which can be calculated by the distance between $L_i(t)$ and $P_j(t)$.

We aim at minimizing the energy cost, which is equivalent to finding the smallest sum of total weights.  A bipartite matching method based on Kuhn-Munkres (KM) algorithm is used. According to KM algorithm, $A[L_i(t)]$ and $B[P_j(t)]$ will be
\begin{align}\label{eq:min_weight}
A[L_i]=&\min\{W[L_i(t)][P_0(t)],W[L_i(t)][P_1(t)],\dots,\nonumber\\
& \qquad \; W[L_i(t)][P_n(t)]\}, \nonumber\\
B[P_j(t)]=&0, \forall i,j\in [1,n].
\end{align}
Since the number of UAV-BSs is equal to the number of predicated positions, a weighted perfect matching can be found using the \emph{Hungarian} algorithm~\cite{doi:10.1002/nav.3800020109}.

\subsection{Proposed Algorithm}
\label{sec:algo}
In this part, the proposed system framework for dynamic placement of multiple UAV-BSs is introduced. Assume $N$ is the number of UEs and $n$ is the number of UAV-BSs. $V_{\textrm{current}}(k)=[x_{\textrm{current}}(k),y_{\textrm{current}}(k)]^T$ is the coordinate of the $k$th UE's current position where $1\leq k\leq N$.

\begin{enumerate}
	\item In the initial time slot, cluster the UEs by using the $K$-means algorithm, and divide UEs into $n$ clusters.
	\item Use the density-aware placement algorithm~\cite{8642333} to find the local optimal positions of each cluster to deploy UAV-BSs in the initial time slot. Let the UAV-BSs fly to those positions. 
	\item Collect the real-time latitude and longitude information of each UE's position every $\lambda$ seconds for $\tau$ seconds.
	\item Use the collected information in 3) as the input UEs' trajectories data for the ESN-based prediction algorithm to predict the UE's positions in the next time slot: $V_{\textrm{predicted}}(k)=[x_{\textrm{predicted}}(k),y_{\textrm{predicted}}(k)]^T$.
	\item Use $V_{\textrm{predicted}}(k)$ as input, calculate the optimal positions of UAV-BSs in the next time slot by using the $K$-means algorithm and density-aware placement algorithm again.
	\item Use the KM-based matching algorithm to find the energy-efficient trajectories of UAV-BSs when moving from current positions to the predicted positions. Let the UAV-BSs fly to those positions.
	\item Repeat from Step 3) to Step 6) after a given time slot $T$. 
\end{enumerate}

In general, the system performance on allocated data rates will be better if $T$ is smaller. However, the total energy cost of UAV-BSs will dramatically increase if $T$ becomes too small. The pseudo-codes of the proposed algorithm are shown in Algorithm~\ref{alg:proposed_algo}.

\vspace{-5pt}
\begin{algorithm2e}[!t]
	\caption{The pseudo-codes of the proposed algorithm}
	\label{alg:proposed_algo}
	\SetAlgoLined	
	\SetKwProg{Fn}{Function}{\string:}{end}
	\SetKwFunction{FPrediction}{Prediction}
	\Fn{\FPrediction{$V(N)$}}{
		Load ESN-based prediction algorithm and update $V(N)\leftarrow [(x_\textrm{predicted}(1), y_\textrm{predicted}(1)),(x_\textrm{predicted}(2), y_\textrm{predicted}(2))$ $,\dots,(x_\textrm{predicted}(N), y_\textrm{predicted}(N))]$\;	
		\KwRet $V(N)$\;
	}
	
	\SetKwFunction{FPlacement}{Placement}
	\Fn{\FPlacement{$L$}}{
		Execute the density-aware placement algorithm and update all the prediction locations of all the UAV-BSs in $L$\;
		\KwRet $L$\;
	}

	\SetKwFunction{FMatch}{Match}
	\Fn{\FMatch{$L,P$}}{
		Load the KM-based matching algorithm to find the minimum weighted perfect bipartite matching $m$ between the current locations $L$ and the predicted locations $P$\;
		\KwRet $m$\;
	}

	\SetKwFunction{FMain}{Main}
	\Fn{\FMain{$N,n,W_N,T,\lambda,\tau$}}{
		Obtain the positions of UEs at the initial time slot $V(N)\leftarrow [(x_\textrm{initial}(1), y_\textrm{initial}(1)),(x_\textrm{initial}(2), y_\textrm{initial}(2))$ $,\dots,(x_\textrm{initial}(N), y_\textrm{initial}(N))]$\;
		Cluster $N$ UEs into $n$ clusters using K-means\;
		At the initial time slot, the initial locations of UAV-BSs are $L=\{L_1,L_2,\dots,L_n\}\leftarrow$ \FPlacement{$L$}\;
		Boolean variable $C=1$\;
		\Repeat{$C\neq 1$}{
			\For{$t=0;$ $t<\tau;$ $t=t+\lambda$}{			
				Collect the real-time positions of all UEs and update $V(N)\leftarrow [(x_\textrm{current}(1), y_\textrm{current}(1)),$ $(x_\textrm{current}(2), y_\textrm{current}(2)),\dots,(x_\textrm{current}(N),$ $y_\textrm{current}(N))]$;
			}	
			$V_\textrm{predicted}(N)\leftarrow$
			 \FPrediction$(V_\textrm{current}(N))$\;
			Cluster UEs into $n$ clusters by using K-means\;
			The prediction result of UAV-BS locations for the next time slot is $P=\{P_1,P_2,\dots,P_n\}\leftarrow$ \FPlacement{$L$}\;
			Do the displacement matching \FMatch{$L,P$}\;
		}
	}

\end{algorithm2e}

\section{Simulation Results}
\label{sec:simulation}
In this section, we discuss the performance of our proposed algorithms. In Section~\ref{sec:4.a}, the Euclidean distance between the predicted trajectory and actual trajectory of the UE is used to evaluate the ESN-based prediction algorithm. In Section~\ref{sec:4.b}, the KM-based matching algorithm is compared to random matching methods in term of the sum of total distance between the current positions and the predicted positions.

\subsection{Performance of Prediction}
\label{sec:4.a}
In the first part, the dataset from the GeoLife project provided by Microsoft Research Asia is used to evaluate our proposed ESN-based prediction algorithm. We select the data in the area of Beijing. 75\% of the data is used as train set and the rest 25\% is used as the test set. In order to adapt the data to the real UAV-BSs application scenario, we set each UE's trajectory data time interval to 3 seconds by interpolating the data and fine-tuning the value of data time.

\begin{figure}[t]
	\centering
	\includegraphics[width=0.5\textwidth]{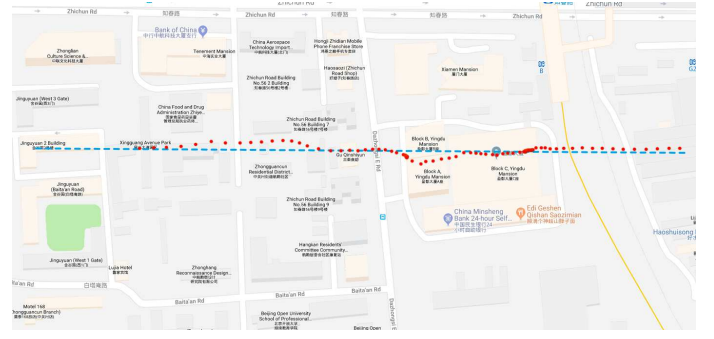}
	\caption{The comparison of predicted and actual trajectories of a UE.}	
	\label{fig:prediction_result}
\end{figure}

The actual trajectory and predicted trajectory of the UE are both depicted with Google Map and shown in Fig.~\ref{fig:prediction_result}. The actual trajectory of the UE is red-dotted line, and the predicted trajectory of UE is blue-dashed line. From this figure, we can find that the predicted trajectory is very close to the actual trajectory.

\begin{figure}[t]
	\vspace{-10pt}
	\centering
	\includegraphics[width=0.5\textwidth]{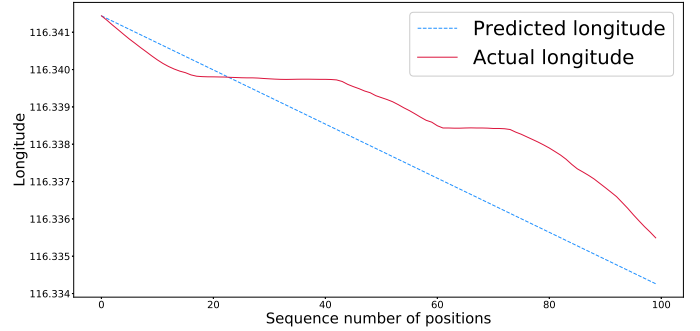}
	\caption{Predicted longitude v.s. actual longitude of a UE.}
	\label{fig:prediction_result:longitude1}
\end{figure}

\begin{figure}[t]
	\vspace{-10pt}
	\centering
	\includegraphics[width=0.5\textwidth]{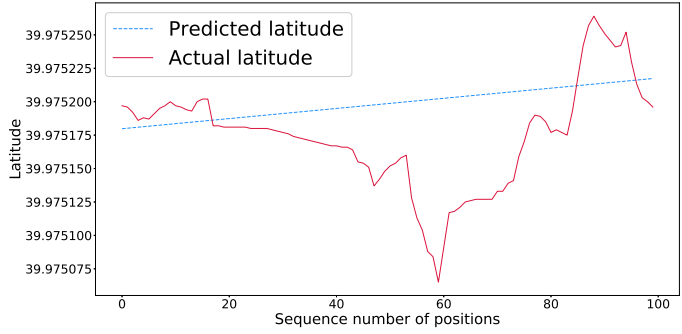}	
	\caption{Predicted latitude v.s. actual latitude of a UE.}
	\label{fig:prediction_result:latitude1}
\end{figure}

We use the actual trajectory data of the UE within 15 minutes as input and predict the trajectory of the UE in the next 5 minutes time slot with the proposed ESN-based prediction model. In Fig.~\ref{fig:prediction_result:longitude1} and Fig.~\ref{fig:prediction_result:latitude1}, the longitude and latitude of predicted trajectory (marked as a blue-dashed line) and actual trajectory (marked as a red-solid line) of the UE are shown respectively. The predicted trajectory and actual trajectory are depicted with 100 predicted positions and 100 actual positions, respectively. The time interval between the predicted positions is set to 3 seconds, which equals to the time interval between the actual positions.

It is obvious that for both longitude and latitude, the predicted value is close to the actual value. Due to the randomness and uncertainty of movement of users, the curve of predicted value is almost smooth, while the curve of actual value fluctuates a little bit. As there is a vertical turn of the actual trajectory in Fig.~\ref{fig:prediction_result:latitude1}, we find the reason after reviewing the actual data, which is the UE changed its direction at that time. We also observe that the curve of predicted longitude and latitude in both Fig. 5 and Fig. 6 are straight. It means that the proposed ESN model cannot predict the trend of the position changes very well. The reason is that there is no sufficient data of non-straight trajectories in GeoLife dataset. 

\begin{figure}[!t]
	\vspace{-10pt}
	\centering
	\includegraphics[width=0.5\textwidth]{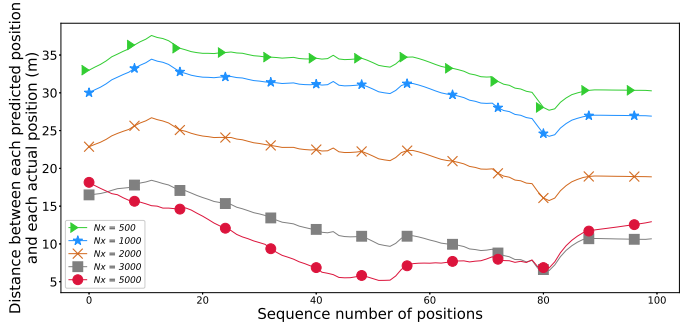}
	\caption{The performance comparisons of the ESN model using different sizes of the reservoir pool.}	
	\label{fig:prediction_result:pool_size}
\end{figure}

\begin{figure}[!t]
	\vspace{-10pt}
	\centering
	\includegraphics[width=0.5\textwidth]{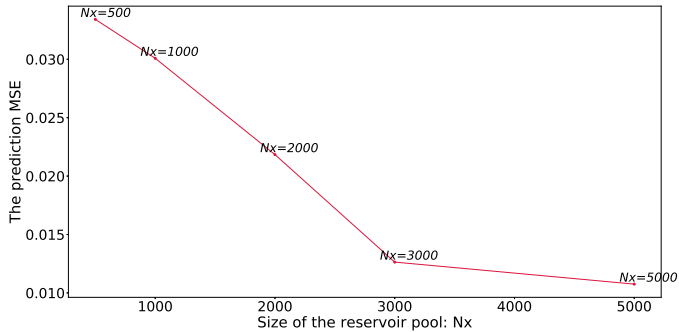}
	\caption{The performance of prediction RMSE.}	
	\label{fig:prediction_result:mse}
\end{figure}

We also evaluate the performance of the ESN model with different sizes of the reservoir pool. We set the size $N_x$ to 500, 1000, 2000, 3000 and 5000, and calculate the distance between the predicted trajectory and actual trajectory of the UE, respectively. According to the result in Fig.~\ref{fig:prediction_result:pool_size}, it indicates that when $N_x$ is set to 5000, the predicted value is closest to the actual value. The RMSE is also calculated and shown in Fig.~\ref{fig:prediction_result:mse}. We find that the RMSE value of each model is no more than 0.030, which indicates that the ESN model has a high prediction accuracy. In addition, when $N_x$ is no more than 5000, the predicted accuracy increases with the increment of $N_x$, which is corresponding to the results in Fig.~\ref{fig:prediction_result:pool_size}.


\subsection{Performance of Matching}
\label{sec:4.b}
We set the number of UAV-BSs to 3. TABLE~\ref{table:latitude_longitude} shows the current latitude and longitude of UAV-BSs, and the predicted latitude and longitude of UAV-BSs in the next 5 minutes time slot. We calculate the distance between each current position and each predicted position when taking the curvature of the Earth into account. The results are shown in TABLE~\ref{table:error_distance}. For instance, if a possible matching scheme $(A\rightarrow\alpha, B\rightarrow\beta, C\rightarrow\gamma)$ is obtained, the UAV-BS whose current position is $A$ needs to fly to the position $\alpha$ in the next time slot. So the distance of the reposition trajectory for this UAV-BS is 267 meters. The rest distance can be deduced in the same way, and the total distance of this matching scheme will be $(267+467+718)=1452$ meters.

We implement the KM-based matching algorithm in Python 3 and use the values in TABLE~\ref{table:error_distance} as the input information. The output reposition matching scheme $(A\rightarrow\gamma, B\rightarrow\alpha, C\rightarrow\beta)$ is shown in TABLE~\ref{table:matching_result}. 

\begin{table}[!t]
	\renewcommand{\arraystretch}{1.2}
	\caption{Current and predicted latitudes and longitudes of UAV-BSs}
	\label{table:latitude_longitude}
	\centering
	\small
	\begin{tabular}{|c|c|c|}
		\hline
		& \textbf{Latitude} & \textbf{Longitude} \\
		\hline
		Current position $A$ & $39.984536$ & $116.316354$\\
		\hline
		Current position $B$ & $39.984501$ & $116.313659$ \\
		\hline
		Current position $C$ & $39.98492$ & $116.314663$ \\
		\hline
		Predicted position $\alpha$ & $39.986506$ & $116.314564$ \\		
		\hline
		Predicted position $\beta$ & $39.988203$ & $116.316238$ \\		
		\hline
		Predicted position $\gamma$ & $39.988461$ & $116.321711$ \\		
		\hline
	\end{tabular}
	\vspace{-5pt}
\end{table}
\begin{table}[!t]
	\renewcommand{\arraystretch}{1.2}
	\caption{The distance between current and predicted positions of UAV-BSs}
	\label{table:error_distance}
	\centering
	\small
	\begin{tabular}{|c|p{1.2cm}|p{1.2cm}|p{1.2cm}|}
		\hline
		& Predicted position $\alpha$ & Predicted position $\beta$ & Predicted position $\gamma$ \\
		\hline
		Current position $A$ & $267$ m & $408$ m & $631$ m\\
		\hline
		Current position $B$ & $236$ m & $467$ m & $851$ m\\
		\hline
		Current position $C$ & $117$ m & $389$ m & $718$ m\\
		\hline
	\end{tabular}
\end{table}
\begin{table}[!t]
	\renewcommand{\arraystretch}{1.2}
	\caption{The reposition matching scheme obtained by KM-based matching algorithm}
	\label{table:matching_result}
	\centering
	\small
	\begin{tabular}{|c|p{1.2cm}|p{1.2cm}|p{1.2cm}|}
		\hline
		& Predicted position $\alpha$ & Predicted position $\beta$ & Predicted position $\gamma$ \\
		\hline
		Current position $A$ & $\times$ & $\times$ & $\bigcirc$ \\
		\hline
		Current position $B$ & $\bigcirc$ & $\times$ & $\times$ \\
		\hline
		Current position $C$ & $\times$ & $\bigcirc$ & $\times$ \\
		\hline
	\end{tabular}
	\\\footnotemark{$\times$: Mismatched    $\bigcirc$: Matched}
\end{table}

Each possible reposition matching schemes and its corresponding sum of distance are listed in TABLE~\ref{table:predicted_sum_distance}. It is obvious that the reposition matching scheme  $(A\rightarrow\gamma, B\rightarrow\alpha, C\rightarrow\beta)$ is the same as the one obtained by the KM-based matching algorithm and it has the minimal sum of the distance. In other words, our KM-based matching algorithm can find the energy-efficient trajectories of multiple UAV-BSs when reposition from the current positions to the predicted positions in the next time slot.
\begin{table}[!t]
	\renewcommand{\arraystretch}{1.2}
	\caption{The distance of all the possible matching schemes}
	\label{table:predicted_sum_distance}
	\centering
	\small
	\begin{tabular}{|p{4cm}|p{3.5cm}|}
		\hline
		Reposition matching scheme & Sum of distance (in meters) \\
		\hline
		$A\rightarrow\alpha, B\rightarrow\beta, C\rightarrow\gamma$ & $1452$ \\
		\hline
		$A\rightarrow\alpha, B\rightarrow\gamma, C\rightarrow\beta$ & $1471$ \\
		\hline
		$A\rightarrow\beta, B\rightarrow\alpha, C\rightarrow\gamma$ & $1362$ \\
		\hline
		$A\rightarrow\beta, B\rightarrow\gamma, C\rightarrow\alpha$ & $1400$ \\
		\hline
		$A\rightarrow\gamma, B\rightarrow\alpha, C\rightarrow\beta$ & $1256$ (minimal)\\
		\hline
		$A\rightarrow\gamma, B\rightarrow\beta, C\rightarrow\alpha$ & $1275$ \\
		\hline
	\end{tabular}
\end{table}

\section{Conclusion}
\label{sec:conclusion}
In this paper, the dynamic placement problem of UAV-BSs is studied. We propose an ESN-based algorithm to predict the future trajectories of the UEs. The predicted trajectories can be used to find the potential positions to reposition the UAV-BSs in the next time slot. Additionally, we consider the energy cost of reposition. A KM-based algorithm is designed to solve the minimum weighted perfect bipartite matching problem and find the minimum energy cost reposition scheme. The simulation results show that the ESN-based algorithm has high accuracy on predicting the next 5 minutes trajectories of the UEs based on previous 15 minutes actual trajectories data, and the matching scheme obtained by the KM-based algorithm satisfies the energy-efficient requirement of UAV-BSs reposition trajectories. In the future, we are going to develop a Generative Adversarial Network (GAN) based framework for generating sufficient trajectory data and thus improve the performance of prediction.

\section*{Acknowledgment}
This research is supported by Ministry of Science and Technology under the Grant MOST 108-2634-F-009-006- through Pervasive Artificial Intelligence Research (PAIR) Labs, Taiwan.



%




\bibliographystyle{IEEEtran}
\bibliography{reference}

\end{document}